\documentstyle[11pt,newpasp,twoside,psfig]{article}
\begin{document}
\def\gtrsim{\mathrel{\hbox{\rlap{\hbox{\lower4pt\hbox{$\sim$}}}\hbox{$>$}}}}
\def\lessim{\mathrel{\hbox{\rlap{\hbox{\lower4pt\hbox{$\sim$}}}\hbox{$<$}}}}

\title{Molecular Gas Reservoir in low-z Powerful Radio Galaxies}
\author{Jeremy Lim}
\affil{Institute of Astronomy \& Astrophysics, Academia Sinica, PO Box 23-141, Taipei 106, Taiwan}
\author{Stephane Leon}
\affil{Instituto de Astrof\'isica de Andaluc\'ia, Granada, Spain}
\author{Fran\b{c}oise Combes}
\affil{Observatoire de Paris, DEMIRM, 61 Avenue de l'Observatoire, 75014 Paris, 
France}
\author{Dinh-V-Trung}
\affil{Institute of Astronomy \& Astrophysics, Academia Sinica, PO Box 23-141, Taipei 106, Taiwan}

\setcounter{page}{111}
\index{de Gaulle, C.}
\index{Churchill, W.}

\begin{abstract}
We report a survey for molecular gas in 3C radio galaxies at redshifts $z \le 0.031$.  Four of the twenty-three galaxies observed were detected with molecular gas masses in the range $10^7$--$10^9 {\rm \ M_{\odot}}$.  The remainder had typical upper limits in molecular gas masses of $\sim$$10^8 {\rm \ M_{\odot}}$.
\end{abstract}

\vspace{-0.9cm}

\section{Introduction}
\vspace{-0.3cm}
Double-lobed extragalactic radio sources are hosted almost exclusively by elliptical galaxies (at low redshifts).  Previous searches for molecular gas in these galaxies have concentrated primarily on those detected in the far-IR, which if produced by dust may indicate copius molecular gas.  A number were indeed found to possess molecular gas with masses typically in the range $\sim$$10^9$--$10^{10} {\rm \ M_{\odot}}$ (see Lim et al. 2000).  Because such molecular gas masses are usually found only in gas-rich disk galaxies and, at masses of $\sim$$10^{10} {\rm \ M_{\odot}}$, merging disk-disk systems, this has prompted the suggestion that radio galaxies comprise the gas-rich E/SO merger products of two gas-rich disk galaxies (e.g., Mazzarella et al. 1983).

To better understand the molecular gas content in the general population of classical (double-lobed) radio galaxies, we have recently completed a CO survey of all previously unobserved or undetected galaxies at $z \le 0.031$ in the revised 3C catalog (Spinrad 1985) with the IRAM 30-m telescope.  Only two of the twenty-five galaxies in this redshift range have previously been detected in molecular gas: 3C~71 (NGC~1068), a relatively nearby ($z = 0.0038$) disk galaxy hosting a compact radio source, and 3C~84 (Perseus~A), a very molecular-gas-rich FR~I elliptical galaxy.  Nineteen of the twenty-three galaxies in our sample are classified as FR~I, two are intermediate FR~I/FR~II, and the remaining three are FR~II galaxies.  Twelve are reported to have been detected in the far-IR, and the same number (although not necessarily the same galaxies) have visible optical dust features.

\section{Results}
\vspace{-0.3cm}
The results of our survey are summarized in Table~1.  Four of the twenty-three galaxies observed were detected, with molecular gas masses spanning two orders of magnitude from $\sim$$1 \times 10^7$ to $\sim$$1 \times 10^9 {\rm \ M_{\odot}}$ (for ${\rm H_o} = 65 {\rm \ Mpc^{-1} \ km \ s^{-1}}$ and ${\rm \Omega_o} = 1$).  Note that the four galaxies detected also have far-IR detections, dust features visible in the optical, and are among those at the lowest redshifts.  The remaining galaxies were not detected at a typical 5$\sigma$ upper limit of $2$--$4 \times 10^8 {\rm \ M_{\odot}}$ (for an assumed linewidth of $\sim$$500 {\rm \ km \ s^{-1}}$).  Our results suggest that only a very small fraction of classical radio galaxies have molecular gas masses $\gtrsim 10^9 {\rm \ M_{\odot}}$, and that the majority contain at most, if any, a few times $10^8 {\rm \ M_{\odot}}$ of molecular gas.

\begin{table}[h]
\caption{Results of CO survey of 3C Radio Galaxies}
\begin{tabular}{lccccc}
\tableline
\tableline
Name     &  FR type  &  Redshift  &  Far-IR  &  Optical Dust  &         $M{\rm(H_2)}$          \\
         &           &    (z)     &          &                &       $(\rm M_{\odot})$        \\
\tableline
3C 31    &    I      &   0.017    &  $\surd$ &    $\surd$     &  $(1.06 \pm 0.04) \times 10^9$ \\
3C 40    &    I      &   0.018    &          &    $\surd$     &  $< 2.3 \times 10^8$           \\
3C 66B   &    I      &   0.021    &          &                &  $< 2.8 \times 10^8$           \\
3C 75N   &    I      &   0.023    &          &                &  $< 3.0 \times 10^8$           \\
3C 75S   &    I      &   0.023    &          &                &  $< 3.0 \times 10^8$           \\
3C 78    &    I      &   0.029    &  $\surd$ &                &  $< 4.4 \times 10^8$           \\
3C 83.1  &    I      &   0.025    &          &    $\surd$     &  $< 4.1 \times 10^8$           \\
3C 88    &    II     &   0.030    &  $\surd$ &    $\surd$     &  $< 3.7 \times 10^8$           \\
3C 98    &    II     &   0.030    &          &                &  $< 4.3 \times 10^8$           \\
3C 129   &    I      &   0.021    &          &                &  $< 1.9 \times 10^8$           \\
3C 129.1 &    I      &   0.022    &          &                &  $< 3.0 \times 10^8$           \\
3C 264   &    I      &   0.022    &  $\surd$ &    $\surd$     &  $(3.09 \pm 0.26) \times 10^8$ \\
3C 270   &    I      &   0.007    &  $\surd$ &    $\surd$     &  $< 3.6 \times 10^7$           \\
3C 272.1 &    I      &   0.003    &  $\surd$ &    $\surd$     &  $(1.40 \pm 0.38) \times 10^7$ \\
3C 274   &    I      &   0.007    &  $\surd$ &                &  standing wave ripple          \\
3C 296   &    I      &   0.024    &  $\surd$ &    $\surd$     &  $< 2.7 \times 10^8$           \\
3C 338   &    I      &   0.030    &  $\surd$ &    $\surd$     &  $< 3.2 \times 10^8$           \\
3C 353   &    II     &   0.031    &          &   tentative    &  $< 4.0 \times 10^8$           \\
3C 386   &    I      &   0.018    &          &                &  $< 2.2 \times 10^8$           \\
3C 402   &    I      &   0.025    &  $\surd$ &                &  $< 3.8 \times 10^8$           \\
3C 442   &   I/II    &   0.026    &          &                &  $< 2.6 \times 10^8$           \\
3C 449   &    I      &   0.018    & tentative &   $\surd$     &  $(2.85 \pm 0.23) \times 10^8$ \\
3C 465   &   I/II    &   0.030    &  $\surd$ &    $\surd$     &  $< 3.9 \times 10^8$           \\
\tableline
\end{tabular}
\end{table}

\end{document}